\documentclass[a4paper,11pt]{article}
\pdfoutput=1 
\usepackage{jcappub} 
\usepackage[T1]{fontenc} 
\usepackage{orcidlink} 

\newcommand{\mpch}{h^{-1} {\rm Mpc}}
\newcommand{\mpcht}{h^{-3} {\rm Mpc^3}}
\newcommand{\gpch}{h^{-1} {\rm Gpc}}

\newcommand{\hmpc}{h {\rm Mpc}^{-1}}

\newcommand{\Rmnum}[1]{\uppercase\expandafter{\romannumeral #1}}
\newcommand{\Msun}{M_{\odot}}

\newcommand{\Kun}{\textsc{Kun} }
\newcommand{\csstemu}{\texttt{CSST Emulator} }
\newcommand{\dlss}{d_\mathrm{LSS}}
\newcommand{\mcal}{\mathcal{M}}

\title{\boldmath Extending CSST Emulator to post-DESI era}
\author[a,b,c,d]{Zhao Chen\orcidlink{0000-0002-2183-9863}}
\author[b,c,d,1]{and Yu Yu\orcidlink{0000-0002-9359-7170}\note{Corresponding author.}}

\affiliation[a]{Tsung-Dao Lee Institute, Shanghai Jiao Tong University, Shanghai 200240, China}
\affiliation[b]{Department of Astronomy, School of Physics and Astronomy, Shanghai Jiao Tong University, Shanghai 200240, China}
\affiliation[c]{State Key Laboratory of Dark Matter Physics, School of Physics and Astronomy, Shanghai Jiao Tong University, Shanghai 200240, China}
\affiliation[d]{Key Laboratory for Particle Astrophysics and Cosmology (MOE)/Shanghai Key Laboratory for Particle Physics and Cosmology, Shanghai 200240, China}

\emailAdd{chyiru@sjtu.edu.cn}
\emailAdd{yuyu22@sjtu.edu.cn}

\abstract{
The recent DESI BAO measurements have revealed a potential deviation from a cosmological constant, suggesting a dynamic nature of dark energy.
To rigorously test this result, complementary probes such as weak gravitational lensing are crucial, demanding highly accurate and efficient predictions of the nonlinear matter power spectrum within the $w_0w_a$CDM framework.
However, most existing emulators fail to cover the full parameter posterior from DESI DR2+CMB constraints in the $w_0\mbox{-}w_a$ plane.
In this work, we extend the spectral equivalence method outlined in Casarini et al. 2016~\cite{2016JCAP...08..008C} to use auxiliary $w_0w_a$CDM models for approximating the power spectrum of a target $w_0w_a$CDM cosmology, moving beyond the previous use of $w$CDM auxiliaries.
Incorporating this enhanced module, the extended \texttt{CSST Emulator} achieves a prediction accuracy of $\leq1\%$ over the $1\sigma$ confidence region from DESI DR2+CMB constraints for $z\leq3$, {with a mild degradation in accuracy outside this posterior region.}
This performance is rigorously validated by additional simulations of dynamic dark energy cosmologies.
The emulator's applicable parameter space has been generalized to fully encompass the $2\sigma$ region, greatly enhancing its utility for cosmological analysis in the post-DESI era.
}

\begin{document}

\maketitle
\flushbottom

\section{Introduction}
\label{sec:intro}

The $\Lambda$CDM model has been the basic cosmological paradigm since the triumph of the precision cosmic microwave background (CMB) observations~\cite{2022ARNPS..72....1T}.
This standard model is spatially flat and consists of two principal components.
There is about $30\%$ pressureless matter, including cold dark matter (CDM) and baryons.
The remaining $\sim70\%$ is the cosmological constant $\Lambda$, also known as the dark energy.
Under Einstein's general relativity, this constant is equivalent to a fluid with the equation of state parameter $w = -1$, providing the negative pressure and driving the acceleration of the cosmic expansion at present (e.g.,~\cite{1998AJ....116.1009R,1999ApJ...517..565P}).
Despite this simple physical scenario, $\Lambda$CDM has met a large number of cosmic observations over the past several decades (e.g.,~\cite{1998AJ....116.1009R,1999ApJ...517..565P,2002MNRAS.337.1068P,2018PhRvD..98d3526A,2020A&A...641A...6P,2021A&A...646A.140H,2022PhRvD.105b3520A,2022MNRAS.511.5492Z}).

Nevertheless, the fundamental nature of dark energy remains unknown, motivating the investigation of alternatives such as diverse dark energy parameterization models and modified gravity theories (e.g.,~\cite{1988ApJ...325L..17P,1995Natur.377..600O,2003RvMP...75..559P,2019LRR....22....1I}).
Most past observations, including CMB, baryon acoustic oscillations (BAO), and weak gravitational lensing, do not find an obvious discrepancy with the $\Lambda$CDM model (e.g.,~\cite{2017MNRAS.470.2617A,2020A&A...641A...6P,2021A&A...646A.140H}).
For instance, the Planck 2018 results combining BAO and supernova place a tight constraint on a constant dark energy equation of state, $w = -1.028\pm 0.031$ at the $68\%$ confidence level~\cite{2020A&A...641A...6P}.
When generalized to dark energy models using the Chevallier-Polarski-Linder (CPL~\cite{2001IJMPD..10..213C,2003PhRvL..90i1301L}) parameterization, this combined constraint still suggests $w_{0} =-0.957\pm 0.080$ and $w_{a} = -0.29^{+0.32}_{-0.26}$, consistent with the $\Lambda$CDM model.
Here, two parameters $w_{0}$ and $w_{a}$ describe the time-evolving equation of state by
\begin{equation}
  w(a) = w_{0} + (1-a) w_{a}\ ,
  \label{eq:CPL}
\end{equation}
where $a$ is the scale factor.

However, the situation is undergoing a notable shift with the release of the first measurements from the Dark Energy Spectroscopic Instrument (DESI) in 2024 \cite{2025JCAP...02..021A}.
This powerful Stage-IV survey detected a $2.6\sigma$ deviation from the cosmological constant for the combination of DESI and CMB under the $w_{0}w_{a}$CDM framework.
When different SNe datasets are incorporated, this discrepancy becomes $2.5\sigma\mbox{-}3.9\sigma$.
These insights into dark energy's dynamic behavior were confirmed by the second data release result (DESI DR2~\cite{2025arXiv250314738D}), sparking extensive discussions on the result and its interpretation.
Consequently, it is now imperative to seriously consider dynamical dark energy models in future cosmological analyses.

Other ongoing and upcoming Stage-IV surveys, such as the Vera Rubin Observatory Legacy Survey of Space and Time (LSST\footnote{\url{http://www.lsst.org}} \cite{2009arXiv0912.0201L}), the Euclid satellite\footnote{\url{https://www.euclid-ec.org}}~\cite{2011arXiv1110.3193L,2024arXiv240513491E}, the Nancy Grace Roman Space Telescope (Roman\footnote{\url{https://roman.gsfc.nasa.gov/}} \cite{2019BAAS...51c.341D}), and the Chinese Space Station Survey Telescope (CSST\footnote{\url{https://www.nao.cas.cn/csst/}} \cite{2019ApJ...883..203G}) will also obtain unprecedented data to advance our knowledge on the nature of dark energy.
On the theoretical side, the accuracy of prediction tools under $w_{0}w_{a}$CDM models must meet the percent-level requirement to match the observational power (e.g.,~\cite{2018PhRvD..98d3532T,2021A&A...649A.100M,2023MNRAS.522.3766T}).
The nonlinear evolution of the Universe makes accurate predictions of small-scale clustering achievable only through cosmological simulations. 
However, the high computational cost makes it infeasible to carry out simulations for a large number of cosmological models.
To overcome this limitation, emulators are designed to interpolate key statistics from a limited set of simulated cosmologies, enabling rapid and accurate predictions across the entire parameter space (see \cite{2022LRCA....8....1A,2023RPPh...86g6901M} for short reviews).
The most basic statistic to emulate is the matter power spectrum, describing the underlying matter clustering as a function of redshift and scale.
Many emulators have been proposed to predict this statistic under different cosmological models (e.g., \texttt{Coyote}~\cite{2010ApJ...715..104H,2009ApJ...705..156H,2010ApJ...713.1322L,2014ApJ...780..111H}, \texttt{Mira-Titan}~\cite{2017ApJ...847...50L,2023MNRAS.520.3443M}, \texttt{BACCO}~\cite{2021MNRAS.507.5869A}, \texttt{EuclidEmulator}~\cite{2019MNRAS.484.5509E,2021MNRAS.505.2840E}, \texttt{CSST Emulator}~\cite{Chen2025} and \texttt{GoKuEmu}~\cite{2025PhRvD.111h3529Y,2026PhRvL.136f1001Y}).
While most of these emulators can not cover the posterior of the DESI DR2+CMB constraint in the $w_{0}\mbox{-} w_{a}$ plane.

Fortunately, previous studies \cite{2005PhRvD..72f1304L,2007MNRAS.380.1079F,2009JCAP...03..014C,2010JCAP...08..005C} have demonstrated that, for a \textit{target} $w_{0}w_{a}$CDM model, the power spectrum at a specified redshift can be well approximated by that from an \textit{auxiliary} $w$CDM cosmology, as long as the comoving distances to the last scattering surface (LSS) of both models are identical.
Casarini et al. 2016~\cite{2016JCAP...08..008C} (hereafter C16) extended the $w$CDM-only \texttt{Coyote} emulator to $w_0w_a$CDM models through this spectral equivalence method and employed extra simulations with $w_{0}=-0.9,\, w_{a}=-0.8$ to validate the sub-percent level of accuracy loss.
In this work, we first assess the robustness of spectral equivalence by using \Kun simulation suite, which covers a broad eight-dimensional parameter space (including dynamic dark energy and massive neutrinos, $\sum m_{\nu}$).
The power spectra of auxiliary $w$CDM models are predicted by \texttt{CSST Emulator}.
After the above internal validation, additional simulations are generated under nine different cosmologies sampled from the posterior of the DESI DR2+CMB constraint.
Spectral equivalence is then tested using results from these dynamic dark energy cosmologies that differ substantially from $\Lambda$CDM.
This indicates that the applicable range of $w_0$ and $w_a$ of \csstemu can be extended to the DESI DR2+CMB posterior.
Furthermore, we demonstrate that spectral equivalence also holds between one target $w_{0}w_{a}$CDM model and one auxiliary $w_{0}w_{a}$CDM model if the condition of comoving distance is satisfied.
This extension increases the prediction accuracy and further broadens the applicable parameter range of the extended \texttt{CSST Emulator}.

The remainder of this article is organized as follows.
Section~\ref{sec:equivalence} demonstrates the physical interpretation and detailed requirements of the spectral equivalence between two different dark energy models.
In Section~\ref{sec:data}, we describe the simulations employed in this study, including the \Kun simulation suite and the extended dynamic dark energy simulations.
In Section~\ref{sec:simulation-validation}, we validate the spectral equivalence inside the original \csstemu cosmological parameter space, as well as the posterior of the DESI DR2+CMB constraint.
The extended range of the \csstemu in $w_{0}\mbox{-}w_{a}$ plane is illustrated in Section~\ref{sec:extended-era}. 
Finally, in Section~\ref{sec:conclusion}, we summarize the results and discuss the directions for further investigation.

\section{Spectral Equivalence Method}
\label{sec:equivalence}

Numerous works have investigated the impact of dark energy on the total matter power spectrum (e.g.,~\cite{2006MNRAS.366..547M,2007ApJ...665..887M,2012MNRAS.419.1588F,2019MNRAS.488.2121C,2022JCAP...11..041P}).
In an expanding Universe, dark energy influences matter clustering by modulating the Hubble expansion rate, which leads to the distinct growth history of matter density fluctuations.
Therefore, it is natural to connect the effect of dark energy to the linear growth factor.
In 2005, Linder \& White~\cite{2005PhRvD..72f1304L} (hereafter LW05) demonstrated that the nonlinear matter power spectra for different constant $w$ models were close to each other ($\sim 1\mbox{-}2\%$) when the linear growth factors at present and a higher redshift were both matched by adjusting the matter density $\Omega_m$.
In this scenario, an important finding is that the comoving distance to LSS, $\dlss$, is nearly equal for these growth-matched models.
Francis et al. 2007~\cite{2007MNRAS.380.1079F} (hereafter F07) proposed to match the $z=0$ nonlinear power spectra by requiring that both $\dlss$ and the linear clustering amplitude parameter $\sigma_{8}$ are identical between different $w$CDM and $w_{0}w_{a}$CDM models.
Different from LW05, other cosmological parameters, such as matter density $\Omega_m$, baryon density $\Omega_b$, dimensionless Hubble parameter $h$, and spectral index $n_s$, were fixed in this matching procedure.
The accuracy of spectral equivalence can achieve $\sim 1\%$ for $k\lesssim 3\, \mpch$ at $z=0$ but becomes worse for higher redshifts.
In order to obtain sub-percent accuracy at various redshifts $z\geq 0$, Casarini et al. 2009~\cite{2009JCAP...03..014C} and 2010~\cite{2010JCAP...08..005C} extended the matching procedure by seeking one single auxiliary $w$CDM model $\mcal_{eq}$ for each given redshift $z$.
Subsequently, C16 utilized this redshift-dependent spectral equivalence to extend the \texttt{Coyote} emulator to predict the nonlinear matter power spectrum in $w_{0}w_{a}$CDM models.

When searching for $\mcal_{eq}$, we maintain present cosmological parameters $\Omega_m$, $\Omega_b$, $h$ and $n_s$ fixed to those of the target $w_{0}w_{a}$CDM model $\mcal$.
This indicates the same CMB prior of the two models due to the identical $\Omega_b h^2$ and $\Omega_m h^2$ at any redshift.
In our work, the sum of neutrino masses $\sum m_{\nu}$ is also fixed to maintain the neutrino contribution to the matter clustering.
Besides, there are two conditions to ensure the spectral equivalence between a target model and an auxiliary model at a given redshift $z$, as established in C16.

The primary impact of dark energy is on the background expansion of the universe, resulting in either an enlargement or a shrinking of the geometric volume.
We can demand the identical comoving distance from the interested $z$ to the redshift of LSS ($z_{*}$) between different models to maintain the size of the universe.
Thus, the first requirement is to match $\dlss$ between a target model $\mcal$ and an auxiliary model $\mcal_{eq}$ by numerically solving the equation
\begin{equation}
\int_z^{z_{*}} \frac{d z^{\prime}}{H_{\mcal_{e q}}\left(z^{\prime}\right)}=
\int_z^{z_{*}} \frac{d z^{\prime}}{H_{\mcal}\left(z^{\prime}\right)} \ .
\label{eq:dlss}
\end{equation}
Here, Hubble parameter $H(z)$ is defined as
\begin{equation}
H(z) = H_0 \left[ \Omega_\mathrm{cb} (1+z)^3 + \Omega_{\nu}(z) + \Omega_r(1+z)^4+  \Omega_\mathrm{de} (z) \right]^{1/2} \ ,
\label{eq:hubble}
\end{equation}
where $\Omega_\mathrm{cb}$ only includes the matter density of CDM and baryons (`cb') during this study.
$\Omega_{\nu}(z)$ represents the redshift-dependent energy density of massive neutrinos with $\Omega_{\nu} = \sum m_{\nu}/93.14\,h^{2} \mathrm{eV}$ at $z=0$.
$\Omega_r$ is the radiation density at present.
For a flat Universe, the evolving dark energy density $\Omega_{de}(z)$ is given by
\begin{equation}
  \Omega_{de}(z)=\left(1-\Omega_\mathrm{cb}-\Omega_{\nu}-\Omega_r\right) \times \begin{cases}(1+z)^{3\left(1+w_{e q}\right)} & \text { for } \mathcal{M}_{e q} \\ (1+z)^{3\left(1+w_0+w_a\right)} \exp \left[-3 w_a z /(1+z)\right] & \text { for } \mathcal{M}\end{cases} \, .
  \label{eq:omega_de}
\end{equation}
Note that for a given target model $\mcal$, different redshifts correspond to different equivalent models, $\mcal_{eq}$.

\begin{figure}[!tbp]
  \centering 
  \includegraphics[width=0.95\textwidth]{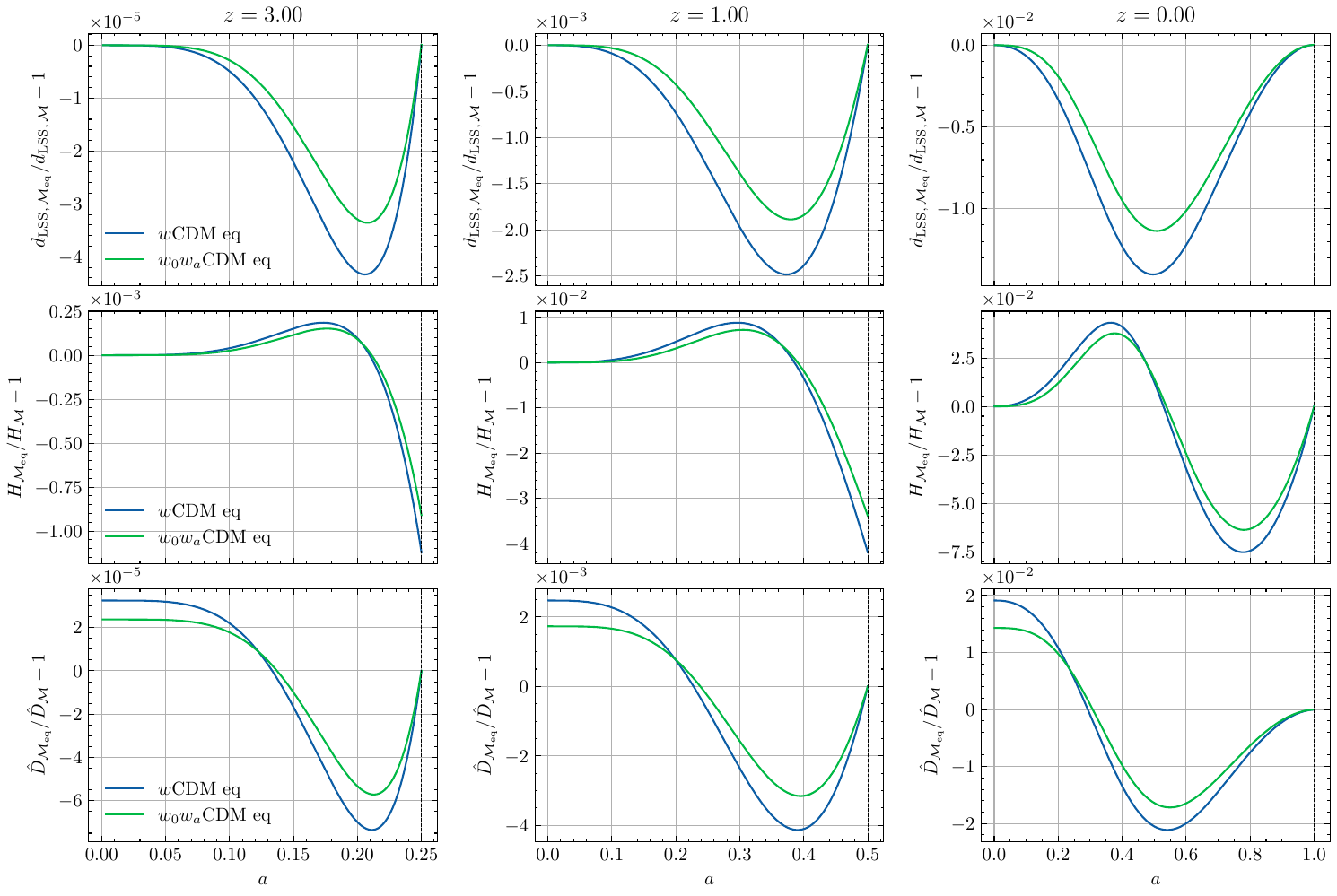}
  \caption{
  {Difference of comoving distance to LSS (top panels), expansion history $H(a)$ (middle panels) and normalized growth factor $\hat{D} = D(a)/D(a=a_{*})$ between a target $w_0w_a$CDM model (dde002) and the corresponding distance-matched models at redshifts $z=$\{3.0, 1.0, 0.0\} (from left to right).
  $a_*$ is the given scale factor at which the distance-matched model is identified, indicated by black dashed lines.
  Blue and green lines represent the distance-matched $w$CDM and $w_0w_a$CDM models (detailed in Section~\ref{sec:desi-era}), respectively.}
  }
  \label{fig:bg-comparison}
\end{figure}

{
To illustrate the consequence of distance matching, we compare the comoving distance to LSS, $\dlss$ (top panels) and the Hubble parameter $H(a)$ (middle panels) between a target $w_0w_a$CDM model and its distance-matched auxiliary models in Figure~\ref{fig:bg-comparison}.
Here we take the dde002 cosmology as an example, which is the cosmology adopted by one of the extended dynamic dark energy simulations described later in Section~\ref{sec:dde-sim}.
Results for other cosmologies are shown in Appendix~\ref{app:bg}.
The black dashed lines in each panel indicate the scale factors $a_*$ at which the distance matching is performed.
Blue and green lines represent the distance-matched $w$CDM and $w_0w_a$CDM models discussed in Section~\ref{sec:desi-era}, respectively.
$\dlss$ of all models are identical at the selected redshifts by construction, while slight deviations are observed at other cosmic times.
These deviations arise from the different expansion histories.
The zero-crossing point of $H_{\mcal_{\rm eq}}/H_{\mcal}-1$ corresponds to the maximum deviation of $\dlss$.
This demonstrates that the target universe initially expands faster than the auxiliary models at high redshifts but becomes slower at low redshifts (or vice versa, depending on the equation of state), leading to an identical physical size of the universe finally.
}

Besides the geometric differences mentioned above, the distinct expansion histories lead to different linear growth factors at lower redshifts.
Consequently, for models that only satisfy the distance-matched condition, the nonlinear matter power spectra still diverge at large scales, even though the initial conditions are identical.
To maintain the identical linear power, we further require the mass fluctuation $\sigma(R, z)$ between $\mcal$ and $\mcal_{eq}$ to be matched.
Here we take $R=8\,\mpch$.
Therefore, the second demand is the amplitude-matched criterion, which is achieved by adjusting the present matter density fluctuation $\sigma_{8, eq}$ for each $\mcal_{eq}$ to satisfy the equation
\begin{equation}
  \sigma_{8, eq}\frac{D_{\mcal_{eq}}(z)}{D_{\mcal_{eq}}(z=0)} = \sigma_8 \frac{D_{\mcal}(z)}{D_{\mcal}(z=0)} \, .
  \label{eq:sigma8}
\end{equation}
In our case, $D(z)$ is the scale-independent linear growth factor derived from Eq.~\ref{eq:hubble}.
Although the linear growth factor of total matter density becomes scale-dependent in the presence of massive neutrinos, the scale-independent growth factor accurately captures the structural growth of CDM and baryons under the Newtonian motion gauge~\cite{2020JCAP...09..018P,2022JCAP...09..068H}.
Several studies have demonstrated that combining the nonlinear `cb' power spectrum with the linear neutrino auto-power spectrum provides a reliable approximation to the nonlinear power for total non-relativistic matter (e.g., \cite{2008PhRvL.100s1301S,2011MNRAS.410.1647A,2014PhRvD..89j3515U,2015JCAP...07..043C,2016ApJ...820..108H}).
Therefore, the method is equally applicable to the clustering of total matter, although the analysis here focuses on the spectral equivalence of the `cb' clustering.
{
The comparison of normalized growth factors $\hat{D} \equiv D(a)/D(a_*)$ is shown in the bottom panels of Figure~\ref{fig:bg-comparison}.
The normalization ensures that $\sigma_8$ is matched at the target redshifts according to Eq.~\ref{eq:sigma8}.
Similar to the comoving distance, the maximum deviation of $\hat{D}$ occurs near the zero-crossing point of $H_{\mcal_{\rm eq}}/H_{\mcal}-1$.}

{
The physical origin of spectral equivalence can be understood as follows.
The nonlinear matter power spectrum at a given redshift is primarily determined by two aspects: (i) the power spectrum shape, which depends on primordial spectrum shape and the expansion history through the horizon scale at matter-radiation equality, and (ii) the nonlinear gravitational evolution, which depends on the linear growth factor and the resulting halo formation history. 
Matching of $\sigma_8$ following Eq.~\ref{eq:sigma8} enforces an identical amplitude for linear density fluctuations.
Supplementary tests in Appendix~\ref{app:inexactdlss} confirm that the distance-matching requirement ensures the auxiliary model’s power spectrum shape aligns with that of the target cosmology, by enforcing identical global cosmic geometry.
Furthermore, these two conditions lead to similar halo formation histories and concentration-mass relations between the target and auxiliary models.
Consequently, key halo model ingredients, including halo abundance, bias, and density profiles, should be approximately preserved, leading to the equivalence of the nonlinear matter power spectrum at extremely small scales ($k\sim 10\,\hmpc$).
However, a comprehensive investigation of halo properties would necessitate large-volume and high-resolution simulations.
We leave this work for future study.
}

{We concluded that at the selected redshifts, both the geometric volume of the universe and the amplitude of matter clustering are matched between the target and auxiliary models, resulting in the spectral
equivalence demonstrated in subsequent sections.
In practice, for a target $w_0w_a$CDM cosmology, we first calculate the target $\dlss$ at a specified redshift.
We then employ the root-finding algorithm to solve for $w_{\rm eq}$ such that the distance-matching condition is satisfied to a relative accuracy of $10^{-7}$.
Once $w_{\rm eq}$ is determined, the $A_{s,\, \mathrm{eq}}$ is adjusted according to the derived $\sigma_{8,\, \mathrm{eq}}$ to satisfy the growth-matching condition in Eq.~\ref{eq:sigma8}.
This procedure uniquely determines the parameters ($w_{\rm eq}$ and $A_{s,\, \mathrm{eq}}$) of the auxiliary model $\mathcal{M}_{\rm eq}$ for each target redshift. 
The computational cost of identifying an equivalent model is negligible ($\sim 1$ minute on a single CPU), relative to $N$-body simulation time, making the method efficient for emulator construction.
}

\section{Data}
\label{sec:data}

Numerical simulations are the most accurate tool to predict the matter power spectrum at small scales.
In this section, we describe the simulations used in this study, including the \Kun simulation suite and the extended dynamic dark energy simulations.

\subsection{\Kun Simulation Suite}
\label{sec:kun-sim}

The \Kun simulation suite\footnote{\url{https://kunsimulation.readthedocs.io/}} is a large set of cosmological $N$-body simulations designed to construct emulators for various statistics~\cite{Chen2025,2025SCPMA..6809513C,2025arXiv250604671Z}, as a part of \textsc{Jiutian} simulations~\cite{2025SCPMA..6809511H}.
This suite covers $129$ different cosmologies under the $w_0w_a\mathrm{CDM}+\sum m_{\nu}$ model, spanning a broad 8D cosmological space.
There are one fiducial cosmology (c0000) taken from Planck 2018~\cite{2020A&A...641A...6P} result and 128 cosmologies (c0001-c0128) sampled through the Sobol sequence sampling method~\cite{sobol1967distribution} over the parameter range
\begin{equation}
  \label{eq:parameter-space}
  \begin{aligned}
  \Omega_{\mathrm{b}} & \in[0.04, 0.06]\ , \\
  \Omega_{\mathrm{cb}} & \in[0.24, 0.40]\ , \\
  n_{\mathrm{s}} & \in[0.92, 1.00]\ , \\
  A_{s} & \in[1.70, 2.50] \times 10^{-9}\ , \\
  H_{0} & \in[60, 80]\ \mathrm{km\, s^{-1}\, Mpc^{-1}}\ , \\
  w_{0} & \in[-1.30, -0.70]\ , \\
  w_{a} & \in[-0.50, 0.50]\ , \\
  \sum m_{\nu} & \in [0.00, 0.30]\ \mathrm{eV}\ .
  \end{aligned}
\end{equation}
Here, $H_0 = 100h\,\mathrm{km\, s^{-1}\, Mpc^{-1}}$ denotes the Hubble constant.
The parameter $A_s$ characterizes the amplitude of the primordial matter power spectrum, which is degenerate with $\sigma_8$.
The influence of massive neutrinos on the `cb' clustering is incorporated utilizing the Newtonian motion gauge method~\cite{2020JCAP...09..018P,2022JCAP...09..068H}.
For each cosmology, we generate a high-resolution simulation with a box size of $L=1\,\gpch$ and $3072^3$ particles.
The particle mass is approximately $2.87 \frac{\Omega_\mathrm{cb}}{0.3} \times 10^{9}\, h^{-1}\Msun$.
All simulations are carried out using the modified Gadget-4 code~\cite{2021MNRAS.506.2871S}.
The initial redshift is fixed at $z_\mathrm{ini}=127$, with the phase of the initial density field preserved across different cosmologies.
Additionally, the fixed amplitude technique~\cite{2016MNRAS.462L...1A} is adopted to suppress the cosmic variance on large scales.
Displacements are computed employing second-order Lagrangian Perturbation Theory (2LPT).
We output 12 particle snapshots at $z=$\{3.00, 2.50, 2.00, 1.75, 1.50, 1.25, 1.00, 0.80, 0.50, 0.25, 0.10, 0.00\}.

In our previous work~\cite{Chen2025}, we successfully constructed the \csstemu to predict the nonlinear matter power spectrum with $\lesssim 1\%$ accuracy for $z\in [0,\,3]$ and $k\leq 10\,\mpch$ under the whole 8D cosmological space shown in Eq.~\ref{eq:parameter-space}.
This outcome is robust, supported by rigorous validation against simulations with even higher mass resolutions (e.g., \textsc{CosmicGrowth} \cite{2019SCPMA..6219511J} and \textsc{AbacusSummit} \cite{2021MNRAS.508.4017M} simulations).
Therefore, it is reliable to validate the spectral equivalence by comparing the predictions of auxiliary $w$CDM models from \csstemu with the original simulation power spectra.
The results are detailed in Section~\ref{sec:csst-emulator-range}.

\subsection{Extended Dynamic Dark Energy Simulations}
\label{sec:dde-sim}

\begin{figure}[!tbp]
  \centering 
  \includegraphics[width=0.6\textwidth]{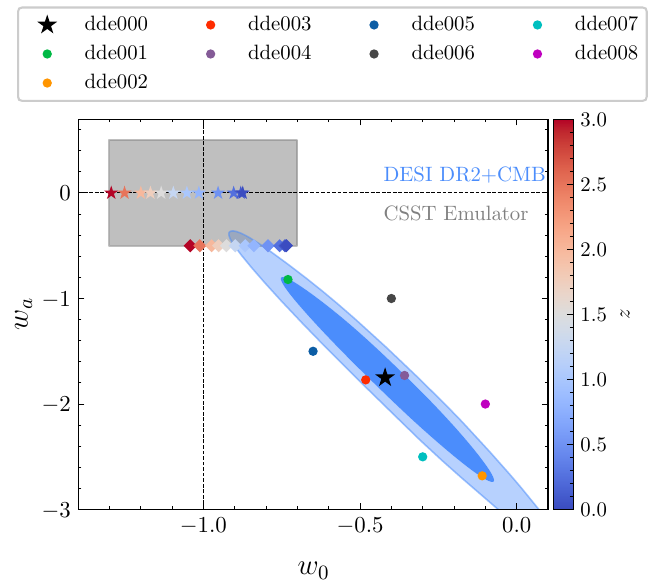}
  \caption{Dark and light blue ellipses represent the $68\%$ and $95\%$ confidence level of DESI DR2+CMB constraint under the $w_0w_a$CDM model. 
  The gray shaded region represents the original parameter range of the \texttt{CSST Emulator}.
  The black star and {eight} colored points denote the cosmologies of the {nine} extended dynamical dark energy simulations used to validate spectral equivalence.
  The colored stars and diamonds indicate the identified auxiliary $w$CDM and $w_0w_a$CDM cosmologies for dde000 cosmology.
  Different colors correspond to redshifts $z\in[0,3]$.
  }
  \label{fig:extended-cosmologies}
\end{figure}

The $68\%$ and $95\%$ confidence levels of DESI DR2+CMB constraint under the $w_0w_a$CDM framework~\cite{2025JCAP...02..021A} are illustrated by the blue ellipses in Figure~\ref{fig:extended-cosmologies}, with the black star indicating the best-fit values.
The \textit{original} \csstemu range of $w_0$ and $w_a$, defined in Section~\ref{eq:parameter-space}, is marked by the gray shaded region, which lies far from the best-fit values.
This motivates running additional simulations to test spectral equivalence within the DESI posterior region.
We set the dde000 cosmology according to the best-fit values of DESI DR2+CMB likelihood with $\Omega_b=0.055$, $\Omega_{m}=\Omega_\mathrm{cb}+\Omega_{\nu}=0.353$, $H_0=63.6\, \mathrm{km\, s^{-1}\, Mpc^{-1}}$, $n_s=0.964$, $\sigma_8=0.781$, $\sum m_{\nu}=0.06\,\mathrm{eV}$, $w_0=-0.42$ and $w_a=-1.75$.
In this cosmology, we perform a high-resolution simulation with the same configuration as the \Kun suite.
{To further support the robustness of the spectral equivalence in the post-DESI era, we take eight other cosmologies from the DESI DR2+CMB posterior, shown by colored points in Figure~\ref{fig:extended-cosmologies}. 
Four of these (dde001-dde004) lie within the $68\%$ confidence region, while the remaining four (dde005-dde008) sample the outer posterior regions.}
These models adopt the same six cosmological parameters as dde000, except for $w_0$ and $w_a$.
For computational efficiency, the simulations of {dde001–dde008} are conducted with $768^3$ particles in a smaller box $250^3\,\mpcht$, achieving the identical mass resolution with the \Kun suite.
To correct the effect of finite simulated volume, we additionally generate a small-box simulation for dde000, keeping the initial phase fixed.
The corresponding $w_0$ and $w_a$ values of the nine cosmologies are summarized in Table~\ref{tab:extended-cosmologies}.

\begin{table}
  \centering
  \begin{tabular}{c@{\hspace{6pt}}*{9}{c@{\hspace{6pt}}}}
    \hline
    Param & dde000 & dde001 & dde002 & dde003 & dde004 & dde005 & dde006 & dde007 & dde008\\
    \hline
    $w_0$ & -0.42 & -0.73 & -0.11 & -0.482 & -0.358  & -0.65 & -0.40 & -0.30 & -0.10\\
    $w_a$ & -1.75 & -0.82 & -2.68 & -1.771 & -1.729 & -1.50 & -1.00 & -2.50 & -2.00\\
    \hline
  \end{tabular}
  \caption{The $w_0$ and $w_a$ values of nine extended dynamic dark energy simulations.
  The other six cosmological parameters are shared with the fiducial cosmology of the extended suite (dde000).
  }
  \label{tab:extended-cosmologies}
\end{table}

\section{Simulation Validation}
\label{sec:simulation-validation}

In this section, we validate the accuracy of spectral equivalence in both the original \csstemu parameter range and the extended DESI DR2+CMB range.

\subsection{Validation inside CSST Emulator Range}
\label{sec:csst-emulator-range}

\begin{figure}[!tbp]
  \centering 
  \includegraphics[width=0.6\textwidth]{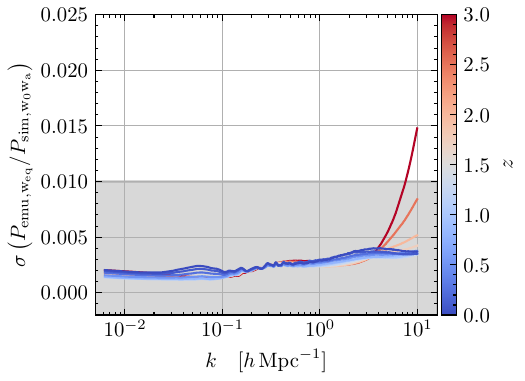}
  \caption{The accuracy of spectral equivalence between the target $w_0w_a$CDM simulations and the auxiliary $w$CDM models predicted by \csstemu at different redshifts.}
  \label{fig:err-pk-csst-range}
\end{figure}

First of all, we utilize $w_0w_a$CDM simulations in \Kun suite to test the robustness of spectral equivalence in the whole 8D cosmological space.
For each redshift $z$ and each target cosmology from c0001 to c0128, we identify one auxiliary $w$CDM cosmology characterized by $w_{eq}$ and $\sigma_{8, eq}$ using the method described in Section~\ref{sec:equivalence}.
Then, we predict the nonlinear `cb' power spectrum of each auxiliary model from \csstemu and compare it with the simulated power spectrum under the target model.
Due to the limited range of $w_0$, only the auxiliary models with $w_{eq}(z) \in [-1.30, -0.70]$ are utilized to validate the accuracy.
Finally, we obtain 104 and 112 validation samples at $z=3$ and $z=0$, respectively.

Figure~\ref{fig:err-pk-csst-range} shows the $68\%$ percentile of the absolute error between the predicted and simulated `cb' power spectra across different redshifts.
Note that the reported absolute error also encompasses the intrinsic systematic error of the emulator.
The accuracy remains better than $0.5\%$ for $z\leq 2$ and slightly degrades to $\sim 1\%$ at higher redshifts.
The modest decline in performance at $z=3$ is probably due to the inherent accuracy limits of \texttt{CSST Emulator}, stemming from the increased contribution of shot noise at high redshifts.
In summary, spectral equivalence proves robust and induces negligible accuracy loss across the broad parameter space defined in Eq.~\ref{eq:parameter-space}, considering the prediction precision of the original emulator.

\subsection{Validation within DESI DR2+CMB Posterior Region}
\label{sec:desi-era}

In this part, we employ extended dynamic dark energy simulations detailed in Section~\ref{sec:dde-sim} to verify the robustness of the spectral equivalence within the posterior region of the DESI DR2+CMB constraint.
Following the approach of the previous section, we seek the auxiliary $w$CDM cosmology for each redshift and target model.
The corresponding $w_{eq}$ values for the dde000 model are illustrated by the colored stars in Figure~\ref{fig:extended-cosmologies}, where different colors represent different redshifts.
All identified $w_{eq}$ values fall entirely within the original \csstemu parameter range, ensuring that the emulator can reliably provide accurate predictions for these auxiliary models.
{However, the derived $w_{eq}$ values exceed the \csstemu range in certain cases.
The auxiliary $w$CDM models for dde002, dde003, dde005 and dde007 yield $w_{eq}<-1.3$ at high redshifts $z \geq 2.5$, while those for dde008 give $w_{eq}>-0.7$ for low redshifts $z \leq 0.1$.
For instance, dde002 and dde007 gives $w_{eq}(z=3.0) = -1.434$ and $-1.532$, respectively.}
Consequently, extrapolations of the \csstemu are employed when predicting the nonlinear spectra for these cases.
Note that for {dde001-009}, only small-volume simulations are available, while for dde000 we have both $1\,\gpch$ and $250\,\mpch$ box simulations. 
Therefore, the ratio of the power spectrum 
from $1\,\gpch$ box simulation to the one from $250\,\mpch$ box simulation is utilized to correct the finite-volume effect in other dynamic dark energy cosmologies.

\begin{figure}[!tbp]
  \centering 
  \includegraphics[width=0.95\textwidth]{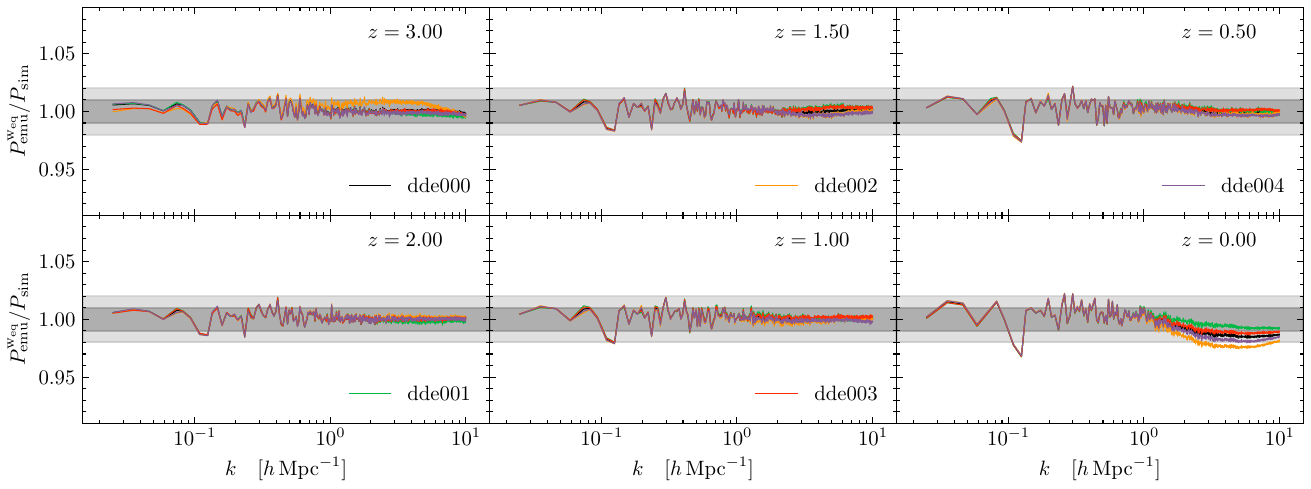}
  \includegraphics[width=0.95\textwidth]{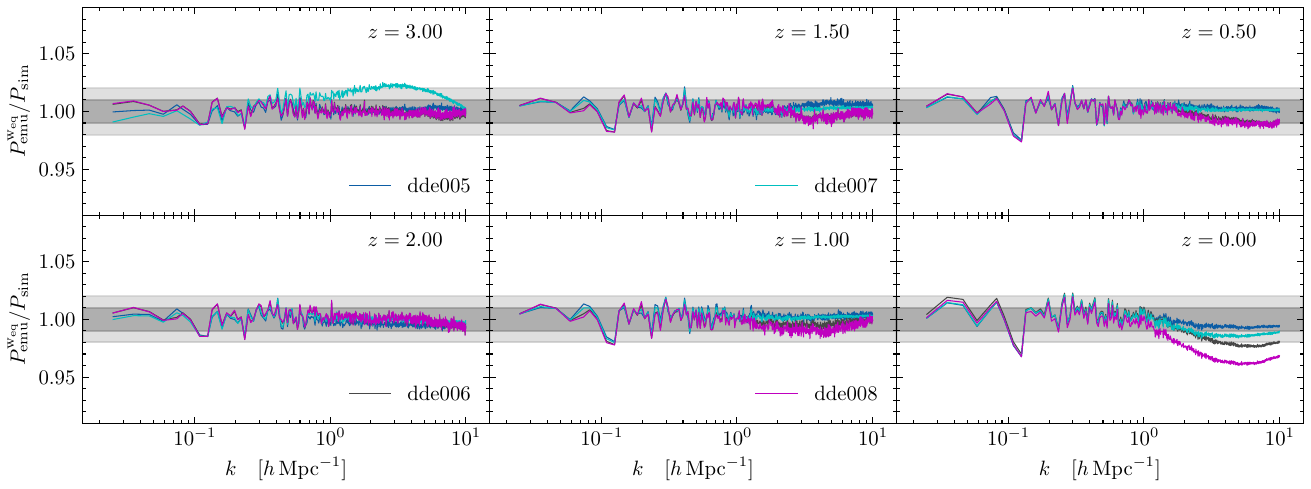} 
  \caption{Comparison of `cb' power spectra between the extended dynamic dark energy simulations and the auxiliary $w$CDM models predicted by \csstemu at different redshifts. 
  Different colors represent different cosmologies.
  The dark and light gray shaded regions indicate the $1\%$ and $2\%$ difference, respectively.}
  \label{fig:weq-emu-vs-sim} 
\end{figure}

The difference of `cb' power spectra between the auxiliary $w$CDM models predicted by \csstemu and the original simulations under nine different $w_0w_a$CDM cosmologies for $0 \leq z \leq 3$ is illustrated in Figure~\ref{fig:weq-emu-vs-sim}.
The dark and light gray shaded regions indicate the $1\%$ and $2\%$ difference, respectively.
{
The overall performance is maintained at $\sim 1\%$ for all redshifts $z\in[0,3]$ and $k\leq 10\,\hmpc$.
At the highest redshift, the slightly higher deviation for dde002 (orange) and dde007 (cyan) is likely due to the degradation in accuracy of the extrapolation.
The deviation marginally exceeds $1\%$ for dde002-dde004 at the most nonlinear region and $z=0$.
While the discrepancy increases moderately for those in the outer posterior region (dde005–dde008).
For the two cosmologies farthest from the best-fit model, dde006 and dde008, the deviations rise to $2\%$ and $4\%$, respectively.
This is reasonable because our matching procedure focuses on the linear growth of structural formation.
A greater distance in the $w_0\mbox{-}w_a$ plane between the target and auxiliary models leads to more significant differences in nonlinear structural growth and, consequently, lower predictive accuracy.
}

To avoid the uncertainty from extrapolation, we need to extend the previous spectral equivalence method.
Inspired by F07, we extend the analysis by considering spectral equivalence between two $w_0w_a$CDM cosmologies, provided the conditions specified in Section~\ref{sec:equivalence} are satisfied.
To this end, both $w_0$ and $w_a$ values are varied in the parameter range of \csstemu to match the comoving distance to LSS when seeking the auxiliary models for dde000-dde004 cosmologies.
Unlike the constant-$w$ case, we can obtain multiple valid solutions within the \csstemu range for each target.
Only the nearest solution to the target $w_0w_a$CDM is selected.
The colored diamonds in Figure~\ref{fig:extended-cosmologies} represent the identified auxiliary $w_0w_a$CDM models for dde000, which cluster near the edge of the gray shaded region.
At each redshift, $\sigma_8$ is then matched to ensure consistent normalization of the linear power spectrum.

\begin{figure}[!tbp]
  \centering 
  \includegraphics[width=0.95\textwidth]{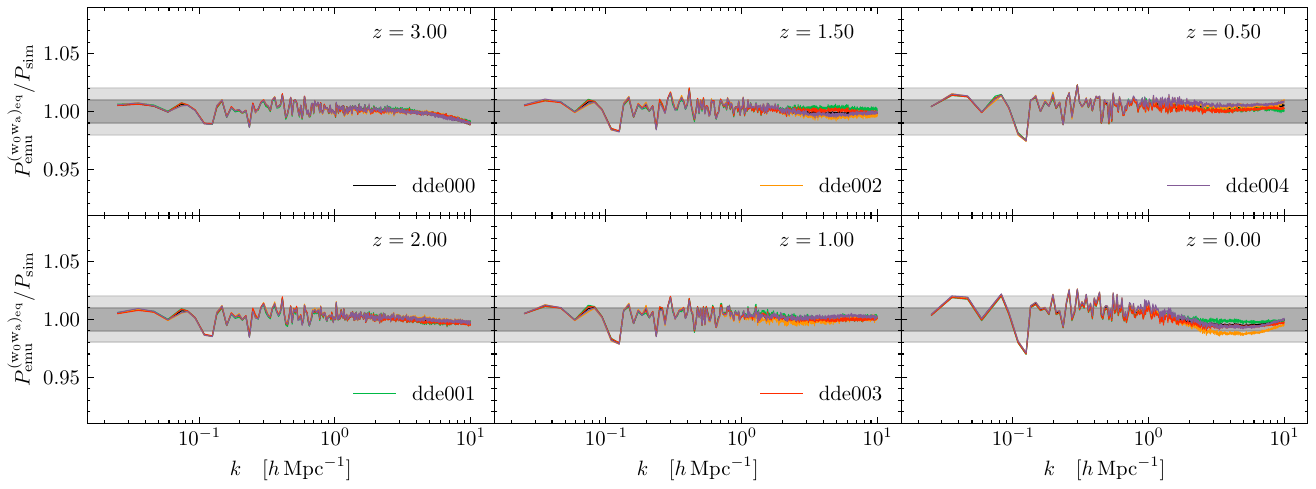}
  \includegraphics[width=0.95\textwidth]{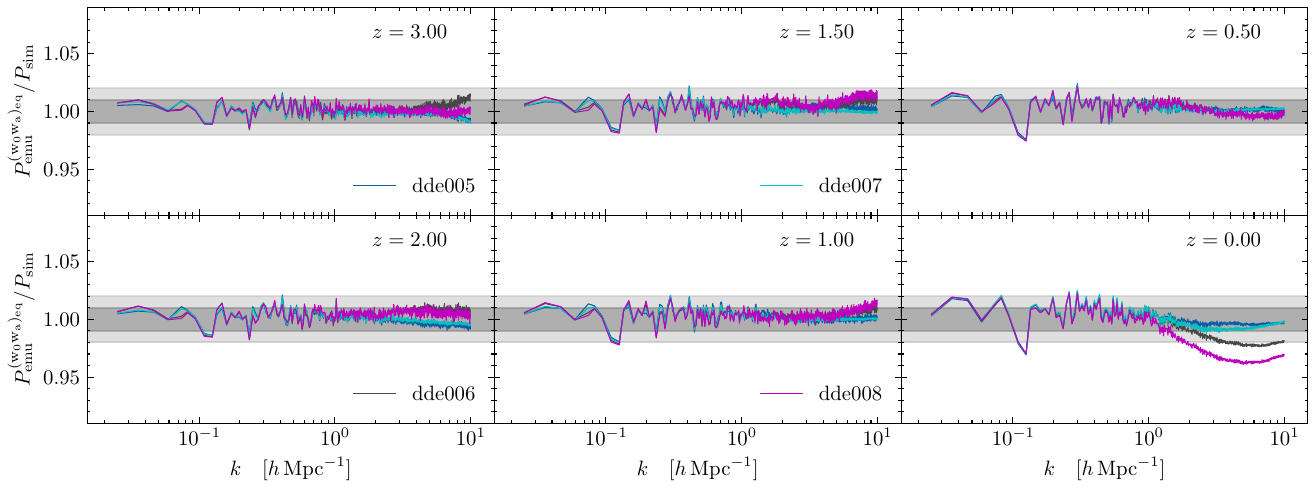} 
  \caption{Similar to Figure~\ref{fig:weq-emu-vs-sim}, but for comparison with the auxiliary $w_0w_a$CDM models.}
  \label{fig:w0wa-emu-vs-sim}
\end{figure}

Figure~\ref{fig:w0wa-emu-vs-sim} presents the comparison between the predicted matter clustering from these auxiliary $w_0w_a$CDM models and the simulation results (including finite-volume corrections) for {dde000–dde009}.
The overall performance is similar to the auxiliary $w$CDM approach (Figure~\ref{fig:weq-emu-vs-sim}), with a significant improvement at the lowest and highest redshifts. 
{
Now for the distant cosmology dde002 and dde007, the accuracy is $\lesssim 1\%$ for $z=0$.
Consistent with the argument in F07, the proximity of two dark energy models in the $w_0\mbox{-}w_a$ plane correlates with reduced spectral differences, explaining the improved agreement.
However, the prediction accuracy of $w_0w_a$CDM models for dde006 (brown) and dde008 (magenta) at $z=0$ has no improvement compared to $w$CDM auxiliary models.
This is caused by the little difference of expansion histories between the $w$CDM and $w_0w_a$CDM auxiliary models for these two cosmologies, detailed in Appendix~\ref{app:bg}.
For $z=3$, all nine dynamic dark energy models achieve highly consistent accuracy at small scales,
as no extrapolation of \csstemu is performed for dde002 (yellow) and dde007 (cyan).}
These results demonstrate that spectral equivalence between two $w_0w_a$CDM models is more robust and can be applied to a broader parameter space.
Therefore, we recommend employing this generalized approach to predict the nonlinear power spectrum beyond the original emulator range.

\subsection{{Comparison with Existing Tools}}
\label{sec:comparison-existing-tools}

{
Considering the unprecedented posterior region of DESI DR2+CMB constraint, most emulators are not designed to cover this parameter space.
Only a few latest emulators, e.g., \texttt{GoKuNEmu}~\cite{2026PhRvL.136f1001Y} and \texttt{Aletheia}~\cite{2025arXiv251113826S} emulator, can be applied to predict the nonlinear power spectrum in this region.
\texttt{GoKuNEmu} emulator is the updated version of \texttt{GoKuEmu}.
Both emulators are constructed based on the \textsc{Goku} simulation suite, which covers a wide parameter space including $w_0\in[-1.3, 0.25]$ and $w_a\in[-3.0, 0.5]$.
While \texttt{GoKuNEmu} achieves $\sim 0.5\%$ average accuracy for $z \in[0,\,3]$ and $k \leq 10 \hmpc$ with the  help of \texttt{T2N-MusE} emulation framework~\cite{2026PhRvD.113d3508Y}.
\texttt{Aletheia} emulator implements the evolution mapping framework using a two-stage Gaussian Process emulation strategy, which shows subpercent accuracy for $0.2<\sigma_{12}<1.0$ and $k< 2\,\mathrm{Mpc}^{-1}$ under the $w_0w_a$CDM model (detailed in Section~\ref{sec:comparison-other-methods}).
However, this emulator does not include the effect of massive neutrinos and thus is not directly comparable to our results.
}

{
Different from emulators, fitting formulae such as Halofit~\cite{2012ApJ...761..152T}, HMcode2020~\cite{2021MNRAS.502.1401M} are based on the halo model, which separately models the contributions from the one-halo and two-halo terms.
The two-halo terms are primarily determined by the linear power spectrum, while the one-halo term is determined by the halo mass function and density profiles, i.e., concentration-mass relation.
If we assume the universality of the halo mass function and concentration-mass relation, it is reasonable to apply these fitting formulae to predict the nonlinear power spectrum in a broader parameter space where the fitted parameters are calibrated.
}

\begin{figure}[!tbp]
  \centering
  \includegraphics[width=0.95\textwidth]{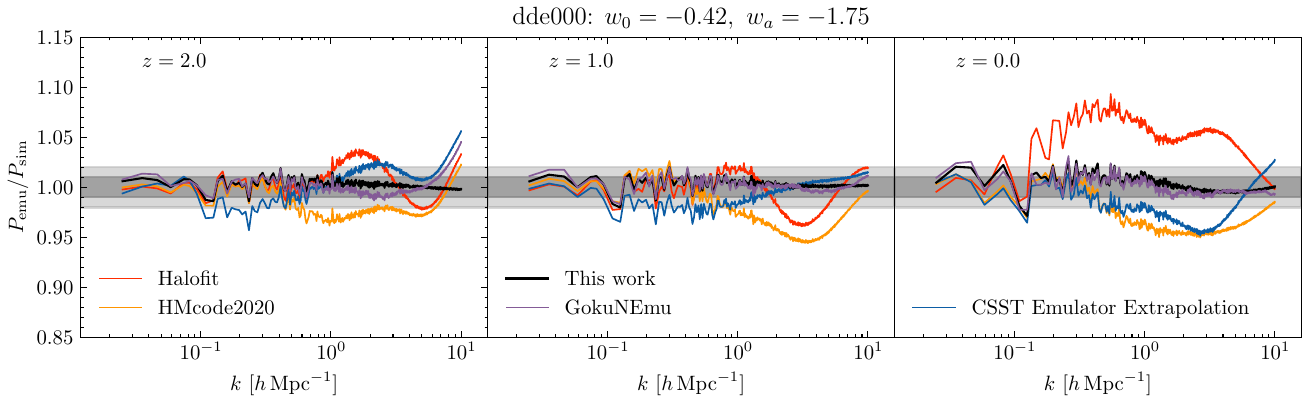}
  \caption{
  {Comparison of `cb' power spectra between the extended dynamic dark energy simulation (dde000) and the predictions from Halofit (red), HMcode2020 (yellow), GoKuNEmu (purple), CSST Emulator with extrapolation (blue) or spectral equivalence (black) at $z = 2.0,\,1.0,\,0.0$. The dark and light gray shaded regions indicate the $1\%$ and $2\%$ difference, respectively.}}
  \label{fig:dde000-pkemu-comparison}
\end{figure}

{
The comparison of nonlinear power spectra between the extended dynamic dark energy simulation (dde000) and the predictions from different tools is shown in Figure~\ref{fig:dde000-pkemu-comparison}.
The red and yellow lines represent the predictions from Halofit and HMcode2020, respectively.
The differences between these two fitting formulae and the simulation results are roughly $5\mbox{-}10\%$.
At the lowest redshift, HMcode2020 performs better than Halofit.
The overall performance is consistent with their claimed accuracy, indicating that the physical parameterization of these fitting formulae can capture the nonlinear evolution of matter clustering to some extent, even for cosmologies with extreme dark energy parameters.
The purple line represents the prediction from \texttt{GoKuNEmu} emulator, which shows excellent agreements ($\lesssim 1\%$) at $z=0$ and $z=1$, and $\sim 3\%$ disagreement only at $z=2.0$ and $k=10\,\hmpc$.
It is likely due to the different simulation setups, e.g., the order of Lagrangian perturbation theory.
All simulations in \Kun utilize 2LPT at $z_\mathrm{ini}=127$, while the \textsc{Goku} simulations adopt the Zel'dovich approximation at $z_\mathrm{ini}=99$.
Combined with the extended spectral equivalence method, \csstemu can achieve $\lesssim 1\%$ accuracy for the nonlinear power spectrum of dde000 at all redshifts and scales.
However, if we directly use the extrapolation results of \csstemu, the accuracy degrades significantly for $k\gtrsim 0.2\,\hmpc$ at all redshifts.
The spectral equivalence method can significantly improve the accuracy of \csstemu predictions for cosmologies beyond the original parameter range.
This is crucial for the analysis of upcoming surveys such as DESI and CSST, as well as for exploring the nature of dark energy.
}

\section{{Discussions}}

\subsection{Extended CPL Parameter Ranges }
\label{sec:extended-era}

Previous results have demonstrated that the spectra of auxiliary $w$CDM or $w_0w_a$CDM models can reliably approximate those of a target $w_0w_a$CDM model if the conditions outlined in Section~\ref{sec:equivalence} are met.
This allows the parameter space of $w_0$ and $w_a$ covered by the original \csstemu to be effectively extended through the spectral equivalence approach.
In this section, we determine the extended parameter space.

\begin{figure}[!tbp]
  \centering 
  \includegraphics[width=0.95\textwidth]{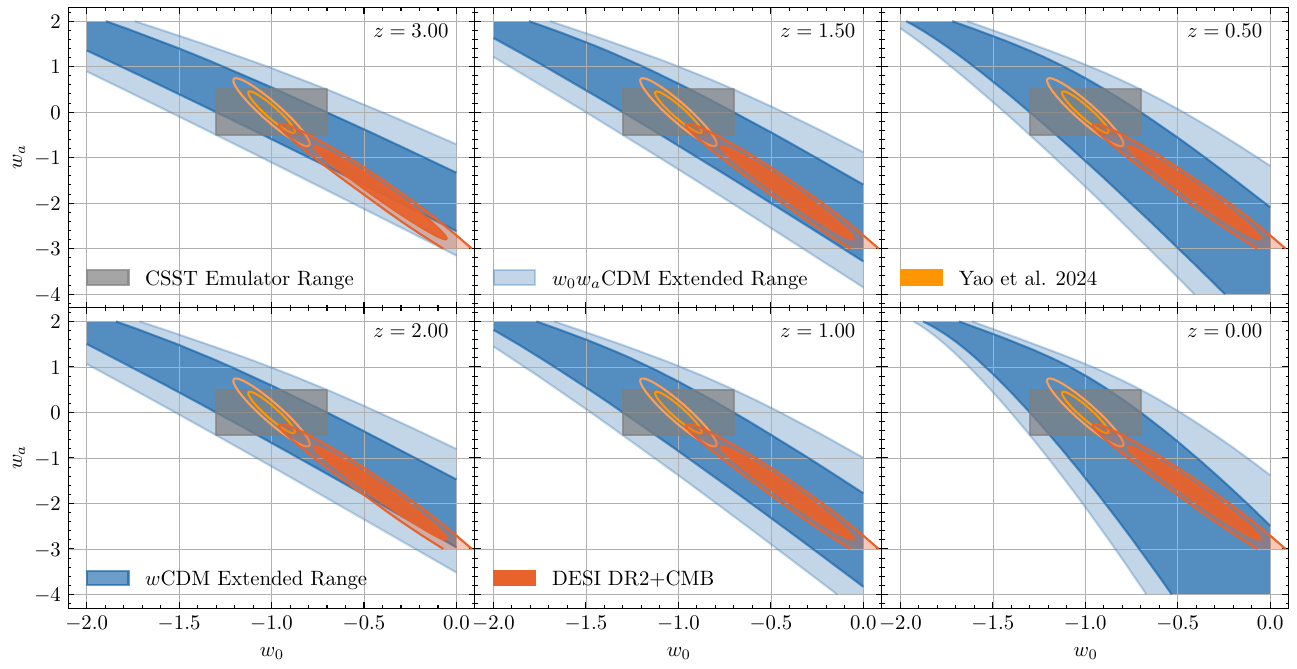}
  \caption{
    The dark and light blue shaded regions represent the extended parameter range of $w_0$ and $w_a$ utilizing $w$CDM and $w_0w_a$CDM as the auxiliary model, respectively.
    Orange ellipses indicate the $68\%$ and $95\%$ confidence level of DESI DR2+CMB constraint under the $w_0w_a$CDM model.
    The original parameter range of \csstemu is shown by the gray shaded region.
    {The forecast of CSST survey is shown by the yellow ellipses, which is derived from the Fisher matrix analysis of the convergence power spectrum $C(\ell)$ with $30\leq\ell \leq 3000$ from Yao et al. 2024~\cite{2024MNRAS.527.5206Y}.}
    Different panels represent different redshifts for $z\in[0,3]$.}
  \label{fig:extended_range}
\end{figure}

In the subsequent analysis, the cosmological parameters $\Omega_m$, $\Omega_b$, $h$, $n_s$, and $\sum m_{\nu}$ are fixed to the values adopted in the extended dynamic dark energy simulations.
A uniform grid of $\sim 30,000$ samples is then generated within the ranges of $w_0\in [-2,0]$ and $w_a\in [-4,2]$.
For each sampled cosmology and redshift, we identify the corresponding auxiliary $w$CDM and $w_0w_a$CDM models as in Section~\ref{sec:desi-era}.
To guarantee the reliability of predictions from \csstemu, we retain only those auxiliary models that satisfy $w_{eq}(z) \in [-1.30, -0.70]$, or $w_{0,eq}(z) \in [-1.30, -0.70]$ and $w_{a,eq}(z) \in [-0.50, 0.50]$.
At each redshift $z\in [0,3]$, the resulting extended ranges of $w_0$ and $w_a$ are illustrated in Figure~\ref{fig:extended_range}.
The dark and light blue regions denote the extended parameter space obtained by seeking auxiliary $w$CDM and $w_0w_a$CDM cosmology, respectively.
While the gray shaded area indicates the original range of \texttt{CSST Emulator}.
The orange ellipses denote the $2\sigma$ contours of the combined DESI DR2 BAO and CMB constraints.
Different panels correspond to different redshifts.
For $z\geq 2.0$, the DESI DR2+CMB posterior slightly exceeds the extended range derived from spectral equivalence with $w$CDM auxiliary models.
This behavior is consistent with the emulator extrapolation observed for the auxiliary models of dde002 and dde007 in Section~\ref{sec:desi-era}.
When the spectral equivalence framework is extended to include $w_0w_a$CDM auxiliary models, however, the entire observational posterior is fully encompassed.
At lower redshifts ($z \leq 1.5$), the extended ranges obtained from both $w$CDM and $w_0w_a$CDM auxiliary models completely cover the $95\%$ confidence region of the DESI DR2+CMB posterior.
{
Furthermore, we also exhibit the forecast of the CSST survey by yellow ellipses, which is derived from the Fisher matrix analysis of the convergence power spectrum $C(\ell)$ with $30\leq\ell \leq 3000$ from Yao et al. 2024~\cite{2024MNRAS.527.5206Y}.
The posterior region of the CSST survey is also fully covered by the extended parameter space.
Although this Fisher forecast is performed using $\Lambda$CDM as the fiducial model, we do not expect the figure of merit to vary much if the DESI best-fit cosmology is chosen as the fiducial.
This demonstrates that the extended \csstemu can provide reliable predictions for the upcoming CSST survey, which will significantly improve the constraints on dark energy properties.
}

The elongated direction of the extended range aligns with the equal-$\dlss$ line in the $w_0\mbox{-}w_a$ plane.
In principle, this direction can be derived by solving the condition of $\dlss = \mathrm{const}$.
However, an analytical solution is not attainable for models with a time-varying dark energy equation of state.
{For practical application, we have incorporated this spectral equivalence method into the \texttt{CSST Emulator}\footnote{\url{https://github.com/czymh/csstemu/blob/master/CEmulator/SpectralEquivalence.py}} package to find the auxiliary models, facilitating cosmological likelihood analysis in the post-DESI era.}
Note that the accuracy degradation over the DESI DR2+CMB posterior region remains negligible relative to the intrinsic precision of the original \texttt{CSST Emulator}, as demonstrated in Section~\ref{sec:simulation-validation}.

{
In practice, the extended $w_0$-$w_a$ range depends on the other six cosmological parameters.
The growth-matching condition requires identical fluctuation amplitudes $\sigma_8(z_*)$ at the target redshift $z_*$.
When $\Omega_m$, $\Omega_b$, $h$, $n_s$, and $\sum m_\nu$ are fixed, $\sigma_8^2(z_*) \propto A_s$.
Users need to adjust $A_s$ to satisfy the growth-matching condition within the \csstemu prior range $A_s \in [1.70, 2.50] \times 10^{-9}$.
We provide the \texttt{SpectralEquivalence\_As} function to compute these adjusted values.
Note that \csstemu accuracy degrades slightly when $A_s$ approaches the lower bound, likely due to increased shot noise contributions (see Appendix A2 of \cite{Chen2025}).
}

\subsection{{Limitations}}
\label{sec:limitation}
{
As discussed in Section~\ref{sec:equivalence}, the spectral equivalence is exact in the linear regime, where the power spectrum evolution is entirely determined by the linear growth factor and the shape of the primordial power spectrum.
In the nonlinear regime, the equivalence becomes approximate.
Nonlinear evolution depends on the full growth history, not just conditions at a single redshift.
Our validation shows that the accuracy remains better than $1\%$ for $k \lesssim 3\,h\,\mathrm{Mpc}^{-1}$ and $z \leq 3$.
At the lowest redshift and smaller scales, accuracy degrades as higher-order nonlinear couplings become important.
The effectiveness of spectral equivalence depends on how closely the auxiliary model approximates the target.
Models that are closer in the $w_0$-$w_a$ plane have more similar expansion histories.
Figure~\ref{fig:bg-comparison} shows that $w_0w_a$CDM auxiliary models track the target expansion history more closely than $w$CDM models.
This explains the improved accuracy in Figure~\ref{fig:w0wa-emu-vs-sim} compared to Figure~\ref{fig:weq-emu-vs-sim}.
At small scales, the nonlinear power spectrum enters the one-halo regime.
Here, clustering is primarily determined by the halo properties, such as the halo mass function and the concentration-mass relation.
Future work could improve accuracy by matching additional quantities, such as integrated structural growth or halo formation history.
In turn, differences in halo properties between target and auxiliary models could help estimate the method's errors.
We leave these investigations for future work.
}

\subsection{{Comparisons with Other Methods}}
\label{sec:comparison-other-methods}

{
In addition to the spectral equivalence method, several other approaches have been proposed and may extend the range of cosmological parameters covered by existing simulations or emulators.
A successful example is the cosmology-rescaling algorithms (e.g.,~\cite{2010MNRAS.405..143A,2019MNRAS.489.5938Z}), providing a complementary framework for reusing simulations across different cosmologies.
It typically requires physical remapping of particle positions, masses and velocities with all cosmological parameters varied simultaneously, achieving $\sim$3\% accuracy for $k\leq 5\,h\,\mathrm{Mpc}^{-1}$ across broad parameter ranges.
However, the accuracy of the cosmology-rescaling method degrades more significantly when the dark energy equation of state deviates from the $\Lambda$CDM model (Figure 6 in~\cite{2019MNRAS.489.5938Z}).
Thus, the constructed BACCO emulator can not cover the DESI DR2+CMB posterior region in the $w_0\mbox{-}w_a$ plane.
In contrast, the spectral equivalence identifies a single auxiliary cosmology at each redshift without post-processing of simulation particles.
This makes spectral equivalence computationally efficient for emulator construction with $\sim 1\%$ accuracy loss for $k\leq 10\,h\,\mathrm{Mpc}^{-1}$.
On the other side, this efficiency comes at the cost of reduced flexibility, as the spectral equivalence is only applicable to cosmologies with the same initial power spectrum shape.
The mapping between the target and auxiliary cosmologies only compensates for differences in the expansion history with other cosmological parameters (e.g., $\Omega_m$, $\Omega_b$, $h$, $n_s$, $\sum m_{\nu}$) fixed.
}

{
Moreover, a conceptually distinct approach is provided by the evolution mapping framework~\cite{2022MNRAS.514.5673S,2024MNRAS.534.3906E,2025PhRvD.112b3520P,2025arXiv251116730F}, which is implemented in the recent \texttt{Aletheia} emulator
\cite{2025arXiv251113826S}.
This method classifies parameters into shape parameters $\mathbf{\Theta}_{\rm s}$, which determine the shape of the linear dimensionless power spectrum $\Delta^2_{\rm L}(k)$ (including $\{\Omega_{\rm b}h^2, \Omega_{\rm m}h^2, n_{\rm s}\}$) and evolution parameters $\mathbf{\Theta}_{\rm e}$ which affect only the amplitude (including $\{h, \sigma_8\}$ and dark energy parameters).
In this framework, the combined effect of evolution parameters and redshift can be compressed into a single amplitude parameter $\sigma_{12}(z)$.
The \texttt{Aletheia} emulator employs a two-stage Gaussian Process emulation strategy. 
The primary emulator predicts the non-linear boost factor as a function of shape parameters $\mathbf{\Theta}_{\rm s}$ and $\sigma_{12}$ for a reference evolution history. 
A second emulator then applies a correction term that is linear in the integrated growth history difference ($\tilde{x}$ in~\cite{2025arXiv251113826S}). 
This linear correction corresponds to a first-order Taylor expansion around the reference evolution. 
The emulator achieves $\sim$0.2\% variance for $w_0w_a$CDM models.
To some extent, the spectral equivalence method can be another expression of the evolution mapping framework.
Both methods achieve sub-percent accuracy and significantly outperform traditional theoretical tools across a wide dark energy parameter range.
}

{
Halo model fitting formulae can predict nonlinear power spectra across broad parameter ranges, as indicated by Figure~\ref{fig:dde000-pkemu-comparison}.
However, this capability relies on the assumption that halo properties, including the mass function and concentration-mass relation, follow universal forms.
Testing this universality for extreme dark energy cosmologies requires additional large-volume simulations.
We defer this investigation to future work.
}

\section{Conclusion}
\label{sec:conclusion}

The unprecedented BAO measurements from DESI have recently revealed indications of a dynamical dark energy component, in contrast to a constant cosmological constant $\Lambda$~\cite{2025JCAP...02..021A,2025arXiv250314738D}.
To rigorously test this exciting result, additional cosmological probes, such as weak gravitational lensing, are required to enhance the precision of parameter constraints.
A key ingredient is the accurate and efficient prediction of the nonlinear matter power spectrum within the $w_0w_a$CDM framework.
However, most existing emulators fail to encompass the DESI DR2+CMB posterior region in the $w_{0}\mbox{-}w_{a}$ plane.
In this study, we employ and extend the spectral equivalence method to enlarge the usable parameter coverage of the accurate \texttt{CSST Emulator}, while preserving negligible loss of precision.
The extended ranges of $w_0$ and $w_a$ at $z\in[0,3]$ can fully enclose the $2\sigma$ confidence contours of the DESI DR2+CMB constraint.

This method has been validated by several previous studies~\cite{2005PhRvD..72f1304L,2007MNRAS.380.1079F,2009JCAP...03..014C,2010JCAP...08..005C,2016JCAP...08..008C}.
Nonetheless, these works were limited by the relatively small number of validation simulations and the omission of massive neutrino effects.
In this work, we exploit the recent \Kun simulation suite to test the robustness of the spectral equivalence across a broad 8D cosmological space, including massive neutrinos.
Inside the original \csstemu parameter range, the discrepancies between predicted spectra of auxiliary models and the corresponding simulation results remain less than $0.5\%$ for $z\leq 2$, increasing only slightly to $\sim 1\%$ at higher redshifts, as shown in Figure~\ref{fig:err-pk-csst-range}.
These results confirm that the accuracy loss due to spectral equivalence is negligible relative to the intrinsic precision of the emulator.

{
To further verify the validity of the spectral equivalence in the post-DESI era, we generate five additional simulations with $w_0$ and $w_a$ values located within the posterior of the DESI DR2+CMB constraint, and four simulations with cosmologies outside the posterior region.
The difference between the predicted `cb' power spectra of auxiliary $w$CDM models from \csstemu and the original simulations under nine different $w_0w_a$CDM cosmologies is illustrated in Figure~\ref{fig:weq-emu-vs-sim}.
The overall accuracy is roughly $1\%$ for all redshifts and scales, while
slightly degrades to $\sim 2-4\%$ at the lowest redshifts for the distant cosmologies.
Moreover, we demonstrate that spectral equivalence between two $w_0w_a$CDM cosmologies outperforms the auxiliary constant-$w$ approach, provided the conditions in Section~\ref{sec:equivalence} are satisfied for both models.
In this case, the spectral accuracy for most cosmologies is moderately improved.
Performance within $1\sigma$ confidence region reaches $\lesssim 1\%$ even in highly nonlinear regimes, as demonstrated in Figure~\ref{fig:w0wa-emu-vs-sim}.
The increased accuracy can be explained by the decreased difference of $H(a)$ between the target cosmology and the auxiliary models in Figure~\ref{fig:bg-comparison}.
In comparison with fitting formulae or existing emulators in Figure~\ref{fig:dde000-pkemu-comparison}, the extended \texttt{CSST Emulator} delivers outstanding predictive performance.
Furthermore, the extended spectral equivalence method significantly broadens the usable parameter space of $w_0$ and $w_a$, fully encompassing the DESI DR2+CMB posterior and the forecast constraint region of the CSST cosmic shear survey at all redshifts $z\in[0,\,3]$, as illustrated in Figure~\ref{fig:extended_range}. 
We conclude that the improved spectral equivalence method in this work enables the \texttt{CSST Emulator} to predict the nonlinear matter power spectrum under the $w_0w_a$CDM framework with both high accuracy and computational efficiency in the post-DESI era.
}

A straightforward extension of this work is to explore the effectiveness of the spectral equivalence method for other statistics, such as the halo mass function, halo clustering, weak lensing peaks, and so on.
While this investigation needs more validation simulations with large volumes, especially for the halo statistics.
We leave it for future work.
Moreover, it is also interesting to explore the optimal criterion for the spectral equivalence method or the physical origin of this phenomenon.
This may further enlarge our knowledge of dark energy effects and even reduce the dimensionality of the dark energy parameter space.
Besides, this spectral equivalence method does not rely on the specific parameterization of dark energy in principle.
Thus, it is worth exploring the robustness of this method for other dark energy models or modified gravity models (e.g., \cite{1974IJTP...10..363H,2000PhRvL..85.4438A,2000PhLB..485..208D,2010PhRvD..81j3528C}). 

\appendix
\section{{Comparisons of Expansion and Growth History}}
\label{app:bg}

\begin{figure}
  \centering 
  \includegraphics[width=0.95\textwidth]{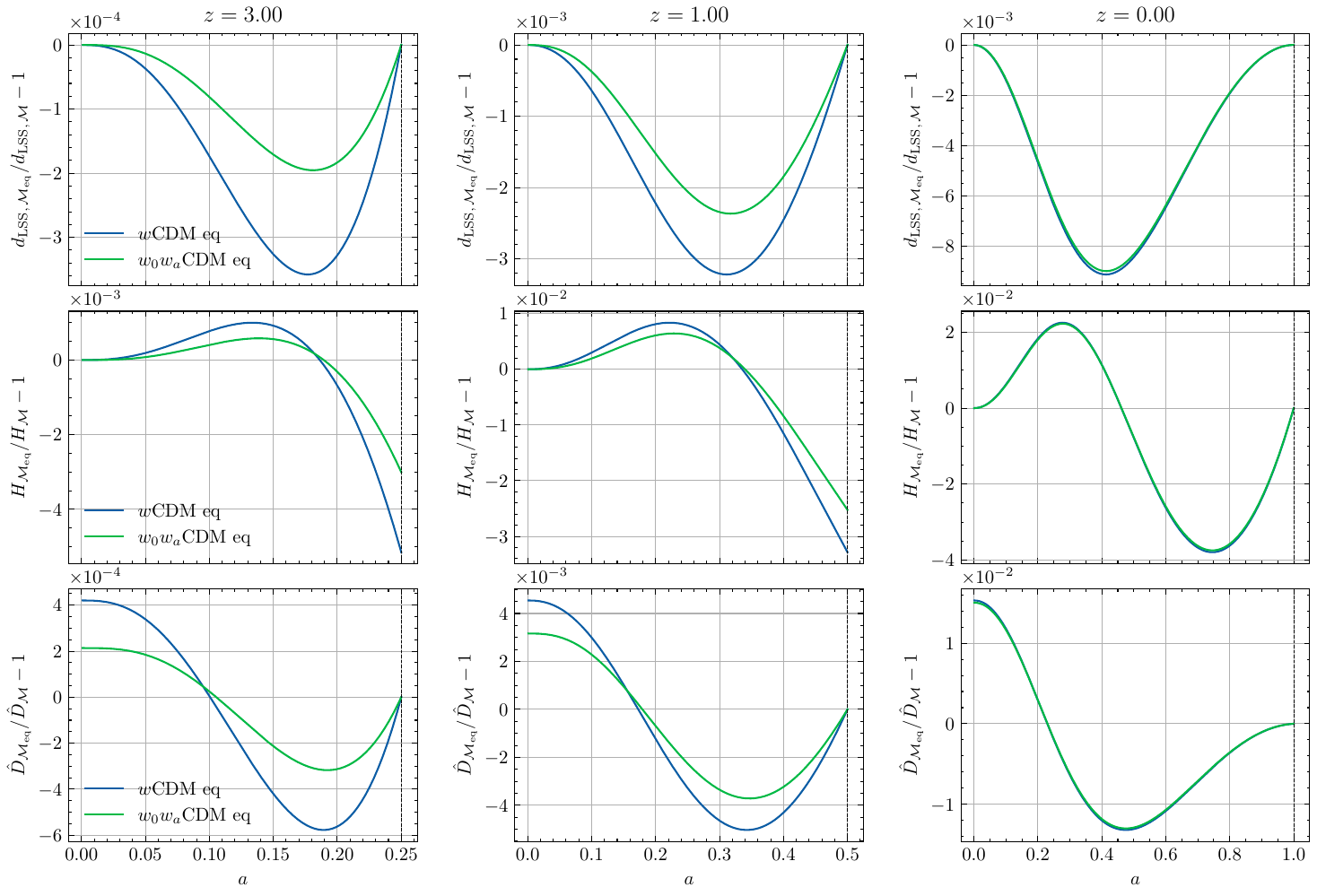}
  \caption{Same as Figure~\ref{fig:bg-comparison}, but for the dde006 cosmology.}
  \label{fig:bg-comparison-dde006}
\end{figure}

\begin{figure}
  \centering 
  \includegraphics[width=0.95\textwidth]{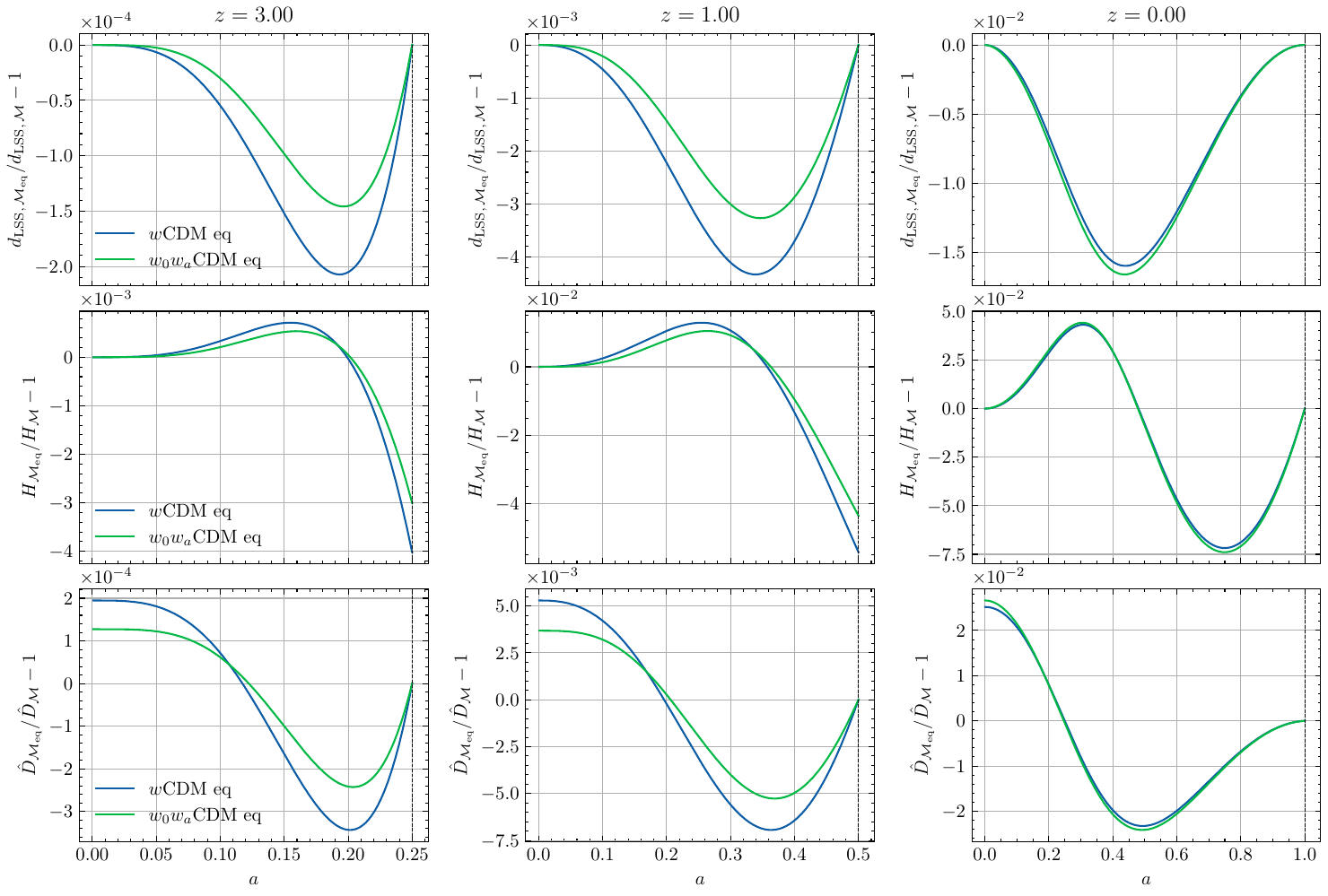}
  \caption{Same as Figure~\ref{fig:bg-comparison}, but for the dde008 cosmology.}
  \label{fig:bg-comparison-dde008}
\end{figure}

{
In this appendix, we present a comparison of the cosmic background evolution between the target $w_0w_a$CDM cosmologies and their auxiliary $w$CDM or $w_0w_a$CDM models, for the set of extended dynamical dark energy simulations introduced earlier.
Results for the dde006 and dde008 cosmologies are shown in Figure~\ref{fig:bg-comparison-dde006} and Figure~\ref{fig:bg-comparison-dde008}, respectively.
In contrast to the dde002 cosmology (displayed in Figure~\ref{fig:bg-comparison}), the discrepancy in cosmic expansion history between the $w$CDM and $w_0w_a$CDM auxiliary models is negligible at $z=0$ for these two cosmologies, leading to the similar accuracy of these two auxiliary models for these two cosmologies in Figure~\ref{fig:weq-emu-vs-sim} and Figure~\ref{fig:w0wa-emu-vs-sim}.
Such close alignment of their cosmic expansion histories stems from the near-identical dark energy parameters between the auxiliary models.
At $z=0$, the auxiliary $w$CDM model for dde006 gives $w_{\mathrm{eq}}=-0.7038$, and its auxiliary $w_0w_a$CDM model yields $(w_{0}, w_{a})_\mathrm{eq}= (-0.7000, -0.0116)$.
For dde008, the corresponding values are $w_{\mathrm{eq}}=-0.6826$ and $(w_{0}, w_{a})_\mathrm{eq}= (-0.7000, 0.0517)$.
This finding demonstrates that the deviation in the $H(z)$ serves as a reliable indicator of the predictive accuracy of the spectral equivalence method, which is consistent with the argument in F07 that the proximity of two dark energy models in the $w_0\mbox{-}w_a$ plane correlates with reduced spectral discrepancies.
}

\section{{Inexact Distance Matching}}
\label{app:inexactdlss}

\begin{figure}
  \centering 
  \includegraphics[width=0.95\textwidth]{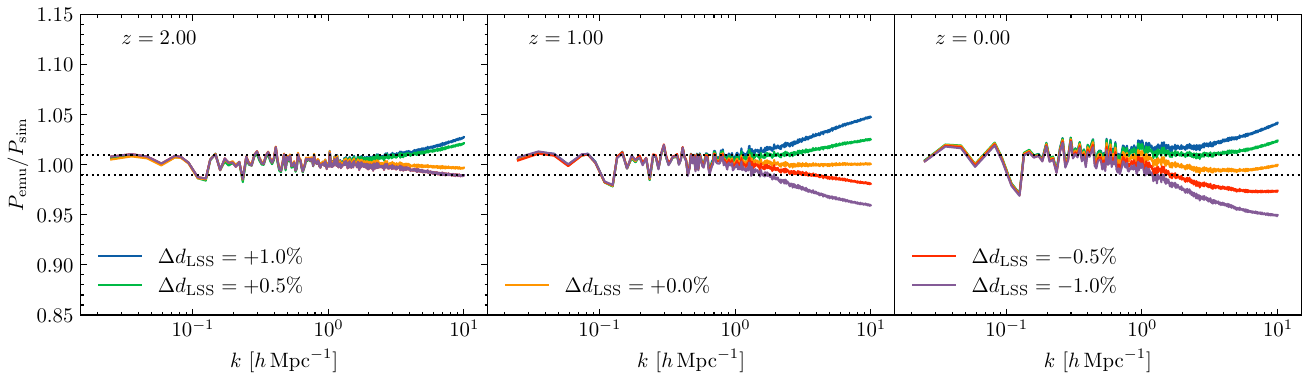}
  \caption{Consequence of the inexact distance-matching condition for dde000 cosmology.}
  \label{fig:inexact-dlss}
\end{figure}

{
The key prerequisite for the spectral equivalence method is the exact matching of the comoving distance to the last scattering surface, $\dlss$, between the target and auxiliary cosmologies.
This condition ensures that the physical scale of the universe is identical at the chosen redshift, which in turn leads to analogous nonlinear structural growth and thus consistent nonlinear matter power spectra.
In this section, we investigate the impact of inexact distance matching on the predictive accuracy of the spectral equivalence method.
To this end, we introduce a small perturbation to the target comoving distance, defined as $\tilde{\dlss}(z) = \dlss(z) \times (1+\Delta \dlss)$, when identifying the corresponding $w_0w_a$CDM auxiliary cosmology.
The resulting deviations in the nonlinear power spectrum are illustrated in Figure~\ref{fig:inexact-dlss}.
This perturbation causes the shape of the auxiliary model’s nonlinear matter power spectrum to diverge from that of the target cosmology.
And the discrepancy is particularly pronounced at small scales.
The accuracy loss is approximately proportional to the amplitude of perturbation, with $\sim 2\%$ accuracy loss at $z=1.0$ and $k=10\,\hmpc$ for $\Delta \dlss \sim 0.5\%$.
Performance degrades more markedly at lower redshifts and smaller scales, where nonlinear evolutionary effects amplify the subtle differences in the cosmic expansion histories of the target and auxiliary cosmologies.
}

\section{{Comparison at Other Cosmologies}}
\label{app:compare-other-cosmos}

\begin{figure}[!tbp]
  \centering
  \includegraphics[width=0.95\textwidth]{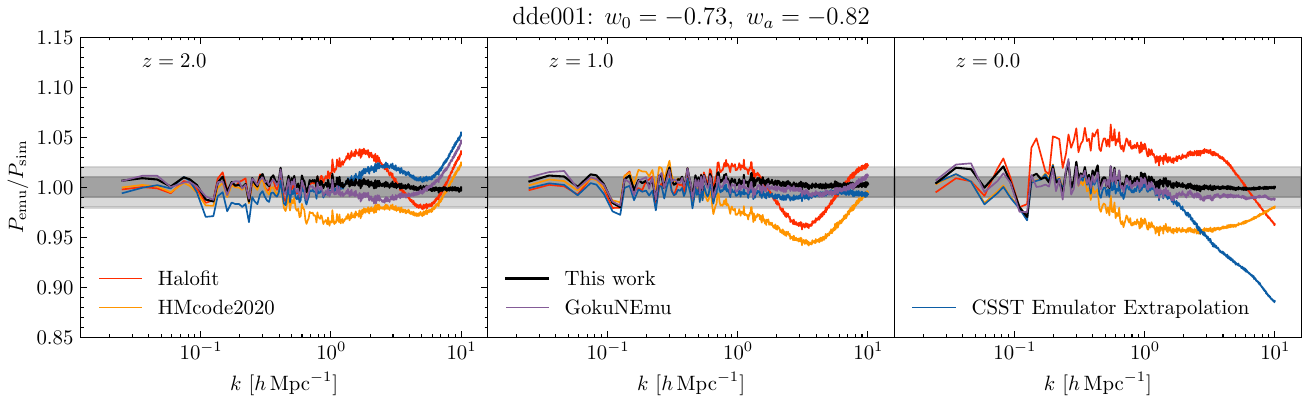}
  \includegraphics[width=0.95\textwidth]{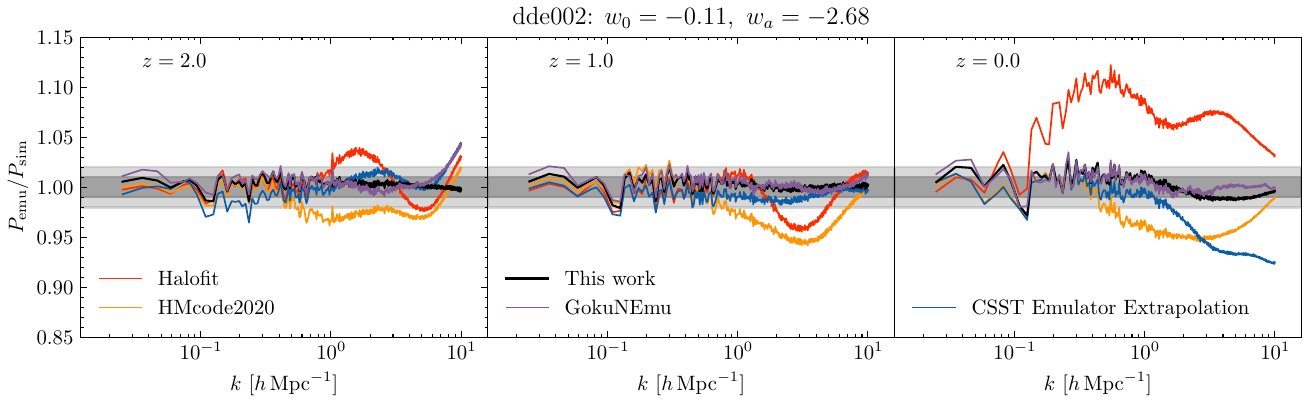}
  \includegraphics[width=0.95\textwidth]{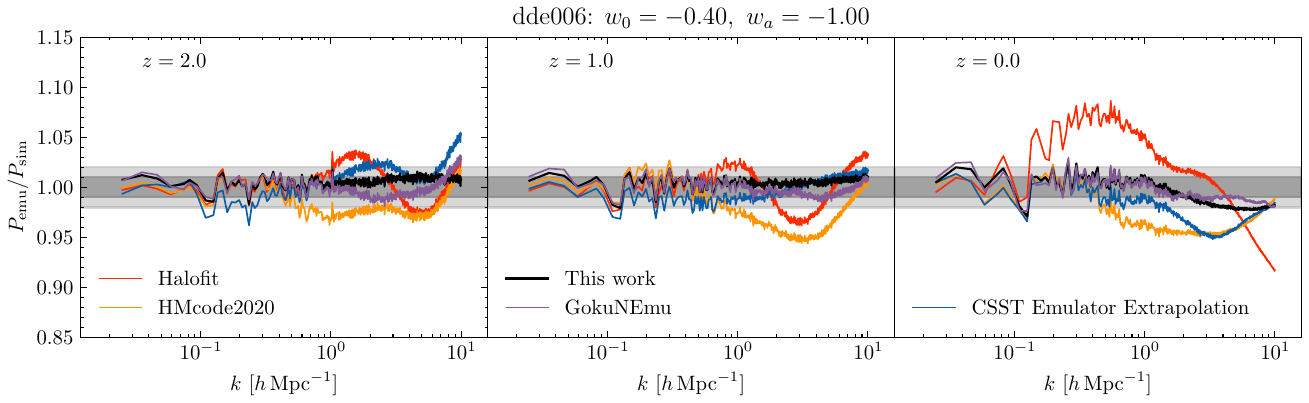}
  \includegraphics[width=0.95\textwidth]{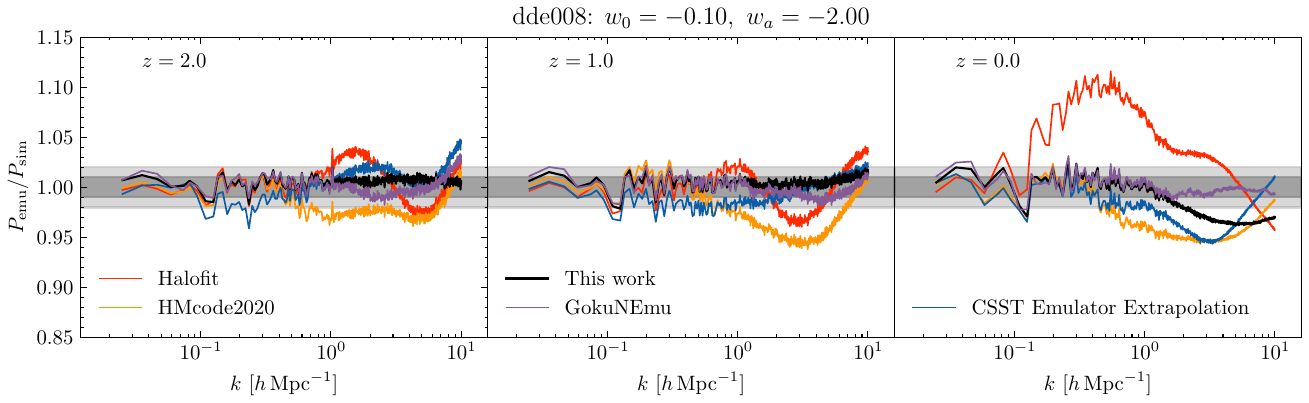}
  \caption{
  Same as Figure~\ref{fig:dde000-pkemu-comparison}, but for dde001, dde002, dde006 and dde008 cosmologies.}
  \label{fig:pkemu-comparison}
\end{figure}

In Figure~\ref{fig:pkemu-comparison}, we compare the spectral equivalence method with Halofit, HMcode2020, GokuNEmu and the direct extrapolation results of \csstemu for dde001, dde002, dde006 and dde008 cosmologies.
Except for the dde008 cosmology, our approach (black) outperforms all other existing tools at all redshifts.
The slightly larger deviation for dde008 at $z=0$ is caused by the larger difference of the expansion histories between the target cosmology and the auxiliary models (detailed in Appendix~\ref{app:bg}). 

\acknowledgments

{We sincerely thank the anonymous referee for the constructive comments, which have rendered our manuscript physically clearer and more readable, and more comprehensive in its entirety.}
This work was supported by the National Key R\&D Program of China (No. 2023YFA1607800, 2023YFA1607801, 2023YFA1607802, 2025YFA1614103), the National Science Foundation of China (Grant Nos. 12595311, 12273020), the China Manned Space Project with Nos. CMS-CSST-2025-A04 \& CMS-CSST-2021-A03, the ``111'' Project of the Ministry of Education under grant No. B20019,
and the sponsorship from Yangyang Development Fund.
This project is supported in part by Office of Science and Technology, Shanghai Municipal Government (grant Nos.\ 
24DX1400100, ZJ2023-ZD-001).
The analysis is performed on the Gravity
Supercomputer at the Department of Astronomy, Shanghai Jiao Tong University.


\bibliographystyle{JHEP}
\bibliography{biblio.bib}


\end{document}